\documentclass[12pt]{article}
\usepackage{color}
\usepackage{amssymb}
\usepackage{epsfig}
\usepackage{footnote}
\usepackage{longtable}
\usepackage{verbatim}
\usepackage{array,multirow}
\usepackage{titlesec}
\usepackage{hyperref}
\usepackage[fleqn]{amsmath} 
\usepackage{graphicx}
\usepackage{caption}
\usepackage{epstopdf}
\include{symbols}
\def\p4D{precision-4D}
\def\P4D{Precision-4D}

\begin{document}
\pagestyle{plain}
%


%
%

\begin{center}
{\Large\bf Cryogenic User Facilities for R\&D on Noble Liquid Detectors and Low Temperature Devices}\\

\vspace{5mm}

{\it A White Paper for Snowmass 2021}\\
\vspace{5mm}

\today
\end{center}

\vspace{5mm}

\begin{center}
Y.~Li, C.~Zhang, M.~Diwan, X.~Qian, S.~Martynenko, C.~Thorn, J.~Stewart, S.~Kettell \\
{\it Brookhaven National Laboratory}\\
\vspace{2mm}
M.~Hollister, A.~Chou, J.~Theilacker \\
{\it Fermi National Accelerator Laboratory}\\
\vspace{2mm}
S.~Golwala\\
{\it California Institute of Technology}\\
\vspace{2mm}
R.~Khatiwada\\
{\it Illinois Institute of Technology}\\
\vspace{2mm}
N.~A.~Kurinsky\\
{\it Stanford Linear Accelerator Center}\\
\vspace{2mm}
C.~Bromberg\\
{\it Michigan State University}\\
\vspace{2mm}
V.~Paolone\\
{\it University of Pittsburgh}
\end{center}
%
%
\begin{abstract}
Cryogenic test facilities are critical infrastructure for physics experiments in a variety of fields, perhaps most notably for particle detection with noble liquid detectors, low-temperature device development, and quantum information research. However, considerable  investment and technical knowledge are required to construct and operate such facilities. This white paper discusses proposals for user facilities aimed at broadening the availability of testing capabilities for the  scientific community.

\end{abstract}
\section{Introduction}
Cryogenic test facilities are critical infrastructure for physics experiments in a variety of fields. Some notable examples include studying noble liquid properties for particle detection, low-temperature device development, and research in quantum information. However, the required technical knowledge and infrastructure capacity, including the cost of setting up and operating the test facilities, can place these studies out of reach of many individual group laboratories. As such, these research areas would greatly benefit from  cryogenic user facilities with suitable access for university and laboratory research groups. This white paper discusses concepts of such facilities, using the liquid argon test facilities and millikelvin test facilities as  examples.

\section{Liquid Argon Test Facilities}

\subsection{Statement of Need}
The Liquid Argon Time Projection Chamber (LArTPC) is an important detector technology that is undergoing rapid development. LArTPCs have been constructed and operated in several neutrino and
dark matter experiments, ranging in size from hundreds of liters to hundreds of cubic meters. Moreover, the Deep Underground Neutrino Experiment (DUNE)~\cite{DUNE} is proposing $\sim$10,000 m$^3$ LAr detector modules to answer important remaining questions in neutrino physics. LAr is chosen because of its ideal physical properties for a detector medium. It is dense and commercially available in large quantities at a relatively low cost. A thorough understanding of LAr properties and optimization of detector designs is crucial to the success of future LAr experiments.

Depending on the specific purpose of the experiment, the active  volume of a test LAr system can vary from $\sim$10 liters to the order of $\sim$10$^3$ liters. In order to accommodate a LArTPC with a fully functional particle detection system, a minimum volume of $\sim$10$^2$ liters is required. Future large LAr detectors will have drift lengths from 0.5  and 6 meters. Studying systematic effects in these large detectors requires a test setup large enough to observe drifts over a substantial distance. In addition to the volume of the cryogenic system, it is equally important to establish the measurement conditions for LArTPC with critical requirements on the purity levels with a quick turnaround time. Impurities in LAr (e.g. oxygen, water, and nitrogen) can significantly attenuate the charge or light signals, resulting in a decrease in detection efficiency and energy resolution. Therefore, it is essential to establish ultra-high purity in the cryogenic system, where considerable care must be taken to purify the Ar and to minimize the introduction of impurities through leakage and surface desorption from materials of the detector systems. 

As a general-purpose LAr test platform, the cryogenic facility needs to have a sufficiently large size to accommodate detector structures, ultra-high purity, excellent control over conditions such as temperature and circulation rate,  operational flexibility with quick turnaround time. Operational flexibility with quick turnaround time implies the ability to stage detectors, or change measurement conditions in the cryostat and be able to start operations without a large overhead in time to get to high purity conditions.


\subsection{Existing Facilities}
Several international institutions have established LAr test stands  for specific purposes or experiments with limited measurement capabilities and limited access to the broader community. Some relatively large-sized facilities (more than $\sim$100 liters of LAr) include the ICARUS 50-liter LArTPC at CERN~\cite{ICARUS:1998nob} with $\sim$250 liter LAr capacity, which is mainly used for ICARUS and DUNE detector's electronics readout studies; the ArgonCube detector~\cite{argoncube} at the University of Bern with $\sim$1100 liter LAr capacity, which is dedicated for pixel readout in LAr and DUNE's modular detector study; and the Integrated Cryostat and Electronics Built for Experimental Research Goals (ICEBERG) at Fermilab with $\sim$3000 liter LAr capability, which is dedicated to DUNE cold electronics studies. As can be seen, these facilities are typically tied to a specific tasks, that are not easily accessible as a user facility with for general purpose R\&D, such as testing new devices and detector designs, studying noble liquid property, calibrating detector signal response, or studying light detector performance.

\subsection{Scope of the Proposed Facility}
We use the 260-liter LAr test platform currently being constructed at the Brookhaven National Laboratory as an example of the proposed general-purpose user facilities. It is an upgrade of an existing 20-liter LAr test stand~\cite{YL_20L}, where the idea of passive argon circulation with gas purification was demonstrated to be effective in achieving the required purity level for LArTPC operation.

The 260-liter main cryostat is intended to host a fully functional TPC to test  detector and electronic readout designs, and measure LAr basic properties for particle detection. The top flange is designed to have sufficient ports for necessary feedthroughs and accommodate future upgrades and applications. An inline liquid argon filter is added to the filling line to achieve parts per billion impurities in terms of oxygen and water upon initial filling, while the gas circulation, purification, and condensing further improves the purity to below 1 ppb within 1 -- 2 days. The turnaround time of the cryogenic operation, in terms of the time starting from cooling down to acquiring the purity level for measurement, can be achieved under one week. 

The necessary initial components for such a LAr test facility include a large cryostat with sufficient volume and drift distance, adequate cryogenic infrastructure for condensing and gas circulation, a purification system to remove impurities with high electron attachment, a purity monitoring system with gas analyzers, basic high voltage system and readout electronics for TPC measurements, a photon detection system to study scintillation light, and the accompanying DAQ system.  
An estimation of the cost to construct and operate such a facility is summarized in Table \ref{tab:cryo}.

\begin{table}[!htp]
\centering
\caption{Approximate unburdened budget 
estimate for a LAr user facility.}
    \label{tab:cryo}
\begin{tabular}{||l|c||} \hline
Line Item & Cost (k\$) \\ \hline \hline
Cryogenic System & 150 \\
Electronics/DAQ system & 50 \\
Optical/Photo Detection system & 50 \\
TPC& 50\\
Infrastructure & 50 \\
Engineering \& Technical Staff & 200 \\
Technical Specialist & 50\\
\hline
\bf{Construction Cost for LAr facility} & \bf{600} \\
\hline \hline
Engineering \& Technical Staff & 100 \\
Cryogens consumption for operation & 50 \\
Materials/Services & 50\\
\hline
\bf{Facility Annual Operations Cost} & \bf{200} \\
\hline
\end{tabular}
\end{table}

\subsection{User Access Model}
With the establishment of the test facility, it is envisioned that general users will apply for access using a simple proposal system stating the purpose, equipment, and time requested for operation. The cost of the user utilizing the facility is expected to be significantly reduced and more efficient compared to the cost of constructing a local test stand. Users can focus more on the fundamental physics measurements and device R\&D without worrying about the complications of operating and maintaining the infrastructure. We expect more innovative technical development and physics results to come out under such a user facility model. The time scale in the next 10--20 years is particularly crucial to support the large upcoming neutrino experiments, such as DUNE, and to realize their full scientific potential.  

\section{Millikelvin Test Facilities}

\subsection{Statement of Need}

The operation of sensors and systems at ultra-low temperature is a major growth field, but one which necessarily has a high barrier for entry due to the relatively high cost of equipment, not just for the cooling platform itself but also associated measurement equipment, electronics, and a knowledge gap in the operation and engineering of devices and systems at cryogenic temperatures.

The establishment of a centralized user  facility would democratize the process of device development
by allowing users access to test stands and measurement equipment for a variety of different
tests, including but not limited to thermal, RF and low frequency tests. This is particularly valuable at the pre-proposal and proof-of-concept stages to validate new ideas before committing to full proposals or the expense of purchasing dedicated test equipment. In addition, such a facility
would serve as a repository of knowledge of low temperature materials and other specialized
information and offer a medium to connect researchers with specialized cryogenic engineering
resources. Finally, such a facility is a vital tool for training a future workforce by providing hands-on experience in the operation of large cryogenic systems and measurement equipment, typically
outside the capability of a University level teaching laboratory.

Such a facility would necessarily consist of more than a single refrigerator test stand, since this is something that a well-equipped PI laboratory would be able to provide. Instead, it is envisaged that the proposed User Facility would provide a suite of test stands available for different experiment types, including a number of test stands providing environments for the testing of specialized devices such underground/low background environments, test stands with optical access, and stands incorporating magnetic fields.

\subsection{Existing Facilities}

While many University and National Laboratory groups operate test stands in the millikelvin temperature region, to the knowledge of the authors none of these test stands are available to the community on a formal and regular  basis.

\subsection{Scope of the Proposed Facility}

An initial set of capabilities for the facility would be as follows:

\begin{itemize}
    \item DC measurements with optional optical access
    \item RF measurements with optional optical access
    \item DC and RF measurements in the presence of a magnetic field
    \item DC and RF measurements in a low-background (underground) environment
    \item Fast-turnaround with minimal DC and RF screening capabilities
    \item Platform for training of operators
\end{itemize}

To satisfy these capabilities, a minimum of 2 large-frame refrigerators (500~mm diameter cold volume) 4-6 medium-frame fridges (with 300~mm diameter cold volume) and 2 small-frame fridges (with 150~mm cold volumes) would be advantageous. 2-3 of the large- and medium-frame refrigerators should be compatible with superconducting magnets. In addition to the test stands themselves, a set of common measurement electronics covering the DC and RF frequency ranges would be required. In particular, modern RF measurement equipment such as network analyzers and signal generators, can easily double the cost of a functional measurement setup. Finally, some scope for infrastructure, including basic vacuum equipment such as pump stations and a leak detector, is included. 


A summary budget for the setup and annual operations of the User Facility is listed in Table \ref{tab:mK}.

\begin{table}[!htp]
\centering
\caption{Approximate budget estimate for a millikelvin user facility}
    \label{tab:mK}
\begin{tabular}{||l|c||} \hline
Line Item & Cost (k\$) \\ \hline \hline
(2) Large frame dilution refrigerators & 2 000 \\
(6) Medium frame dilution refrigerators & 3 000 \\
(2) Small frame dilution refrigerators & 600 \\
Helium-3 for 10 fridges & 200 \\
(3) Superconducting Magnets & 1 500 \\
RF Measurement Equipment & 2 000 \\
DC Measurement Equipment & 500 \\
Infrastructure and site preparation & 700 \\
Engineering \& Technical Staff & 700 \\
\hline
\bf{Setup Cost} & \bf{11 000} \\
\hline \hline
Engineering \& Technical Staff & 500 \\
Clerical \& Administrative Staff & 100 \\
Consumables \& Maintenance & 200 \\
\hline
\bf{Annual Operations Cost} & \bf{800} \\
\hline
\end{tabular}
\end{table}

\subsection{User Access Model}

It is envisaged that a user access model similar to that of existing user facilities would be adopted in which potential users will apply for time using a simple proposal system defining the type of payload and measurement desired. The motivation of the facility is to promote proof-of-concept and pre-proposal projects. It is expected that Users will be required to pay nominal fees for access to the facility in order to support Operations, with rates determined by the affiliation of the User (for example, commercial Users would be required to pay a higher rate for time than an academic User).

\section{Conclusions}
\label{conclusions}

The establishment of a LAr user facility will provide a general purpose test
platform for the community with quick turnaround time to reach desired operation conditions. Small-scale experiments can be performed without a huge setup overhead. There will be many opportunities to use this facility. Examples include but not limited to: measurement of fundamental LAr properties, such as the diffusion coefficient, in an external electric field, measurement of light and charge yield at different stopping powers, testing of new TPC designs and cold electronics readout, testing of new photocathode materials and electron sources, and testing of materials that enhance the detector performance. We believe such a long term LAr test facility will integrate well and be beneficial to the future LAr detector R\&D programs. In a similar manner, the establishment of a millikelvin user facility will fill a broad gap in the community by centralizing testing capabilities and engineering and measurement knowledge, overcoming the considerable financial and technical barrier to PIs operating their own local facilities, serving to democratize the field.  We also expect that such cryogenic user facilities will greatly enhance the collaborations between universities and national laboratories, and facilitate more efficient use of resources. 



\bibliographystyle{unsrt}
\bibliography{main}

\begin{thebibliography}{1}

\bibitem{DUNE}
Babak Abi et~al.
\newblock {Deep Underground Neutrino Experiment (DUNE), Far Detector Technical
  Design Report, Volume I Introduction to DUNE}.
\newblock {\em JINST}, 15(08):T08008, 2020.

\bibitem{ICARUS:1998nob}
F.~Arneodo et~al.
\newblock {The ICARUS 50-l LAr TPC in the CERN neutrino beam}.
\newblock In {\em {INFN Eloisatron Project: 36th Workshop: New Detectors}},
  pages 3--12, 12 1998.

\bibitem{argoncube}
ArgonCube Collaboration.
\newblock {https://argoncube.org/}.

\bibitem{YL_20L}
Yichen Li et~al.
\newblock {A 20-Liter Test Stand with Gas Purification for Liquid Argon
  Research}.
\newblock {\em JINST}, 11(06):T06001, 2016.

\end{thebibliography}






\end{document}